\documentclass[conference]{IEEEtran}
\IEEEoverridecommandlockouts
\usepackage[binary-units]{siunitx}
\usepackage{tabularx}
\usepackage{multirow}
\usepackage{color, colortbl}
\usepackage[binary-units]{siunitx}
\DeclareSIUnit{\bps}{bps}
\usepackage{algorithm2e}
\usepackage{amsmath,amssymb,amsfonts}
\usepackage{cite}
\usepackage{graphicx}

\ifCLASSINFOpdf
\else
\fi
\begin{document}
\title{Opportunistic AP Selection in Cell-Free Massive MIMO-OFDM Systems}

\author{\IEEEauthorblockN{Wei Jiang\IEEEauthorrefmark{1}\IEEEauthorrefmark{2} and Hans D. Schotten\IEEEauthorrefmark{2}\IEEEauthorrefmark{1}}
\IEEEauthorblockA{\IEEEauthorrefmark{1} German Research Center for Artificial Intelligence (DFKI), Trippstadter Str. 122,  Kaiserslautern, 67663 Germany\\
  }
\IEEEauthorblockA{\IEEEauthorrefmark{2}Technische Universit\"at (TU) Kaiserslautern, Building 11, Paul-Ehrlich Street, Kaiserslautern, 67663 Germany\\
 }
 \thanks{This work was supported by the German Federal Ministry of Education and Research (BMBF) through the \emph{Open6G-Hub} project (Grant no.  \emph{16KISK003K}).}
}


%


\maketitle

\begin{abstract}
Exploiting the degree of freedom in the frequency domain and the near-far effect among different access points (APs), this paper proposes an opportunistic transmission scheme in cell-free massive MIMO-OFDM systems. The key idea is to orthogonally assign subcarriers among different users, so that there is only one user on each subcarrier. Then, a user is only served by its near APs through opportunistic selection, while the far APs are deactivated to avoid wasting power over their channels with severe propagation losses. Moreover, the number of active APs per subcarrier becomes small due to the opportunistic selection, making the use of downlink pilots and coherent detection feasible. As corroborated by numerical results, the proposed scheme can bring a significant performance boost in terms of both power efficiency and spectral efficiency.
\end{abstract}


%
\IEEEpeerreviewmaketitle

\section{Introduction}
Cell-free massive multi-input multi-output (CFmMIMO)  has recently attracted a lot of interests from both academia and industry \cite{Ref_ngo2017cellfree}, which is promising to become a key technology in the six-generation mobile system \cite{Ref_jiang2021road}. Until now, several key issues such as energy efficiency \cite{Ref_ngo2018total}, power control \cite{Ref_nayebi2017precoding}, scalability \cite{Ref_bjornson2020scalable}, backhaul constraint \cite{Ref_masoumi2020performance},  channel impairment \cite{Ref_jiang2021impactcellfree}, resource allocation \cite{Ref_buzzi2020usercentric}, and pilot assignment \cite{Ref_zeng2021pilot}, have been reported in the literature. The previous works assume the wireless channels undergo flat fading. Nevertheless, most of wireless systems nowadays are wideband with signal bandwidths far wider than the \textit{coherence bandwidth}, leading to frequency selectivity. To fill this gap, the authors of this paper proposed and analyzed orthogonal frequency-division multiplexing (OFDM)-based multi-carrier transmission for CFmMIMO, coined cell-free massive MIMO-OFDM (CFmMIMO-OFDM) in \cite{Ref_jiang2021cellfree}, over frequency-selective fading channels. As a follow-up aiming to further unleash the potential of CFmMIMO-OFDM, we propose an opportunistic transmission scheme in this paper by exploiting the degree of freedom in the frequency domain and the near-far effect among distributed APs due to the cell-free structure.

The key idea is to assign orthogonal frequency-domain resources to different users so that each subcarrier or resource block (RB) carries only a single user. This setup not only avoids multi-user interference but also simplifies the system design. Then, a number of access points (APs) with strong large-scale fading (defined as the near APs hereinafter) are opportunistically selected to serve this user. At the same time, the far APs with weak large-scale fading are deactivated over the assigned subcarriers or RBs for this user. As a side effect, the number of active APs per subcarrier becomes small, which enables the use of downlink pilots so that the user can perform coherent detection. The major technical benefits of the proposed opportunistic AP selection (OAS) scheme are two-fold:
\begin{itemize}
    \item \textit{Opportunistic Gain}: From the point of view of a typical user, a near AP has a favorable channel with a small path loss. In contrast, the energy radiated from a far AP is wasted over its long propagation distance. In other words, the same amount of power transmitted from a near AP generates a much stronger received power than a far AP, resulting in high power and spectral efficiencies.
    \item \textit{Coherent Gain}: From the perspective of each sub-carrier or resource block, only a few APs serve a single user. Then, a high-dimensional massive MIMO system is transformed into a low-dimensional MISO system. As a result, the prohibitive overhead of inserting downlink pilots, which is proportional to the massive number of base-station antennas, can be alleviated. The user can obtain the instantaneous channel state information (CSI) by estimating downlink pilots, rather than only knowing statistical CSI, and then perform coherent detection. Hence, a fundamental problem restricting the downlink performance of massive MIMO can be solved thanks to opportunistic AP selection.
\end{itemize}

The remainder of this paper is organized as follows: the next section introduces the system model of CFmMIMO-OFDM. Section III and IV present the OAS scheme and analyze its performance, respectively. Section V explains the simulation setup and shows some numerical results, followed by the conclusions in Section VI.

\section{System Model}
Consider a CFmMIMO system where $M$ randomly distributed APs that are connected to a central processing unit (CPU) serve $K$ users in a geographical area. Without losing generality, we assume that each AP and user equipment (UE) is equipped with a single antenna as \cite{Ref_ngo2017cellfree} for simple analysis. The results are applicable to multi-antenna APs with an extension, which is out of the scope of this paper. The frequency-selective fading channel between AP $m$ and user $k$  can be modeled as a linear time-varying filter in a baseband equivalent basis \cite{Ref_tse2005fundamental}, i.e.,
\begin{equation}
    \mathbf{h}_{mk}[t] = \Bigl[ h_{mk,0}[t],\ldots, h_{mk,L_{mk}-1}[t]  \Bigr]^T,
\end{equation}
where the filter length $L_{mk}$ depends on the delay spread and the sampling interval.
Taking into account large-scale fading $\beta_{mk}$\footnote{We observe that the cell-free structure results in the \textit{near-far} effect among different APs from the perspective of a typical user. The APs can be therefore divided into two categories: the near APs and the far APs, similar to the near and far users from the perspective of a base station in the conventional cellular systems.}, the channel filter between AP $m$ and user $k$ can be modelled by
\begin{align} \nonumber
\mathbf{g}_{mk}[t]&=\Bigl[ g_{mk,0}[t],\ldots, g_{mk,L_{mk}-1}[t]  \Bigr] ^T\\
&=\sqrt{\beta_{mk}} \mathbf{h}_{mk}[t],
\end{align}
with $g_{mk,l}[t]=\sqrt{\beta_{mk}} h_{mk,l}[t]$, $\forall l=0,1,\ldots,L_{mk}-1$.

The OFDM transmission is \cite{Ref_jiang2016ofdm} organized in block-wise.
We denote the frequency-domain symbol block of AP $m$ on the $t^{th}$ OFDM symbol by $\tilde{\mathbf{x}}_m[t] = \left[ \tilde{x}_{m,0}[t],\ldots, \tilde{x}_{m,N-1}[t]  \right]^T$. Performing an $N$-point inverse discrete Fourier transform (IDFT), $\tilde{\mathbf{x}}_m[t]$ is converted into a time-domain sequence $\mathbf{x}_m[t] = \left[ x_{m,0}[t],\ldots, x_{m,N-1}[t] \right]^T$ in terms of
\begin{equation}
    x_{m, n'}[t]=\frac{1}{N}\sum_{n=0}^{N-1}\tilde{x}_{m, n}[t] e^{\frac{2\pi jn'n}{N}},\:\:\:\: \forall n'.
\end{equation}
Defining the discrete Fourier transform (DFT) matrix
\begin{equation}
\label {Eqn_DFTMatrix}
\mathbf{D} =
\left[ \begin{aligned}
         \Omega_N^{00} &  &  \cdots && \Omega_N^{0(N-1)} \\
         \vdots &&  \ddots && \vdots \\
         \Omega_N^{(N-1)0} &  &  \cdots && \Omega_N^{(N-1)(N-1)}
\end{aligned} \right]
\end{equation}
with a primitive $N^{th}$ root of unity  $\Omega_N^{n n'}=e^{-\frac{2\pi jn'n}{N}}$, the OFDM modulation is expressed in matrix form as
\begin{equation} \label{eqn:transmitsignal}
   \mathbf{x}_m[t] =\mathbf{D}^{-1} \tilde{\mathbf{x}}_m[t]=\frac{1}{N}\mathbf{D}^{*}\tilde{\mathbf{x}}_m[t].
\end{equation}

Cyclic prefix (CP) is inserted between two transmission blocks to preserve subcarrier orthogonality and absorb inter-symbol interference. The transmitted baseband signal with CP is denoted by $\mathbf{x}_m^{cp}[t]$.
Going through the wireless channel, it results in a received signal component $ \textbf{x}_m^{cp}[t] \ast \mathbf{g}_{mk}[t]$ at the typical user $k$, where $\ast$ denotes \textit{the linear convolution}. Consequently, the received signal at user $k$ is given by
\begin{equation}
    \mathbf{y}_k^{cp}[t]=\sum_{m=1}^M  \mathbf{g}_{mk}[t]\ast \textbf{x}_m^{cp}[t] +\mathbf{z}_k[t],
\end{equation}
where $\mathbf{z}_k[t]$ denotes a vector of additive white Gaussian noise with zero mean and variance $\sigma_z^2$, i.e., $\mathbf{z}_k\sim \mathcal{CN}(\mathbf{0},\mathbf{\sigma_z^2\mathbf{I}})$.
Removing the CP, we get
\begin{equation} \label{eqn:reveivedsignal}
  \mathbf{y}_k[t]=\sum_{m=1}^M  \mathbf{g}_{mk}^N[t]\otimes \mathbf{x}_m[t] +\mathbf{z}_k[t],
\end{equation}
where $\otimes$ stands for \textit{the cyclic convolution}, and $\mathbf{g}_{mk}^N[t]$ is an $N$-length zero-padded vector of $\mathbf{g}_{mk}[t]$. Then, the frequency-domain received signal is computed by
\begin{equation} \label{eqn:frequencydomainRx}
    \tilde{\mathbf{y}}_k[t] = \mathbf{D}\mathbf{y}_k[t].
\end{equation}
Substituting (\ref{eqn:reveivedsignal}) into (\ref{eqn:frequencydomainRx}), and applying \textit{the convolution theorem} \cite{Ref_jiang2016ofdm}, yields
\begin{IEEEeqnarray}{lll}
\label{Eqn_conditionedx} \nonumber
\tilde{\mathbf{y}}_k[t] &=& \sum_{m=1}^{M} \mathbf{D}\left(\mathbf{g}_{mk}^N[t]\otimes\mathbf{x}_m[t]\right)+\mathbf{D}\mathbf{z}_k[t]\\
         & =& \sum_{m=1}^{M} \tilde{\mathbf{g}}_{mk}[t] \odot \tilde{\mathbf{x}}_m[t] + \tilde{\mathbf{z}}_k[t],
\end{IEEEeqnarray}
where $ \odot$ represents the Hadamard product.
For a typical subcarrier $n$, the downlink signal model is expressed by
\begin{equation}
\label{Eqn_OFDMDL}
   \tilde{y}_{k,n}[t]=\sum_{m=1}^M \tilde{g}_{mk, n}[t]\tilde{x}_{m,n}[t]+\tilde{z}_{k,n}[t], \:\:k\in\{1,\ldots,K\}.
\end{equation}

\begin{figure}[!bpht]
\centering
\includegraphics[width=0.325\textwidth]{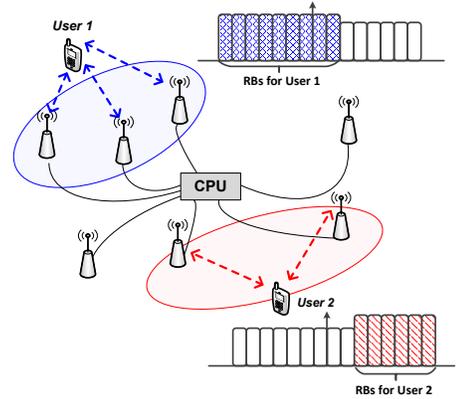}
\caption{Illustration of opportunistic AP selection in a cell-free massive MIMO-OFDM system. }
\label{Diagram_CFM}
\end{figure}

\section{Opportunistic AP Selection}
The downlink transmission from the APs to the users and the uplink transmission from  the users to the APs are separated by time-division multiplexing (TDD) with the assumption of perfect channel reciprocity. The length of a radio frame is generally less than the channel coherence time, and therefore the channel condition is regarded as constant within a frame. Without losing generality, we only focus on the downlink transmission in this paper but the application of the proposed scheme in the uplink is straightforward. We ignore the time index of signals for simple analysis hereinafter.

The communication process of the OAS scheme in a cell-free massive MIMO-OFDM system is depicted as follows:
\begin{itemize}
    \item AP $m$, $\forall m$ measures the large-scale fading $\beta_{mk}$, $k=1,2,\ldots,K$, in a long-term basis, and reports this information periodically to the CPU. Thus, the CPU has a global knowledge of large-scale CSI $\mathbf{B}\in \mathbb{C}^{M\times K}$, where $\left[\mathbf{B}\right]_{m,k}=\beta_{mk}$ with $[\cdot]_{m,k}$ denotes the $(m,k)^{th}$ entry of a matrix. Since $\beta_{mk}$ is frequency independent and varies slowly, this measurement is practically easy to implement.
    \item UE $k$, $\forall k$ periodically reports its data-rate request with a scalar $r_{q,k}$ through the uplink signalling. Then, the CPU knows $\mathbf{r}_{q}=\{r_{q,1},r_{q,2},\ldots,r_{q,K}\}$.
    \item \textit{Frequency-Domain Resource Allocation:} According to some particular criteria, e.g., fairness, priority, and performance, the CPU makes a decision of resource allocation as a function of the users' requests, i.e.,  $\{\mathbb{B}_1,\ldots,\mathbb{B}_K\}=f(\mathbf{r}_{q})$, where the specific implementation of $f(\cdot)$ is out of the scope of this paper. The resource pool consists of $N$ OFDM subcarriers, denoted by the set of subcarrier indices $\mathbb{B}=\left\{0,1,2\ldots,N-1 \right\}$. Using $\mathbb{B}_k$ to denote the indices of subcarriers assigned to user $k$, we have $\bigcup_{k=1}^K \mathbb{B}_k \in \mathbb{B}$ (when all subcarrier are allocated, $\bigcup_{k=1}^K \mathbb{B}_k = \mathbb{B}$). The subcarriers are allocated orthogonally, satisfying $\mathbb{B}_k \bigcap \mathbb{B}_{k'}=\varnothing$, $\forall k'\neq k$. The time interval of resource allocation depends on the system design.
    \item \textit{Opportunistic Selection:} The CPU selects the opportunistic APs for each user in terms of large-scale fading. Assume the number of selected APs is $M_s$, where $1\leqslant M_s\leqslant M$. Order the indices of the APs in terms of their large-scale fading in a descending order, and then select the first $M_s$ APs. We denote the set of opportunistic APs for user $k$ by $\mathbb{M}_k=\{\ddot{m}_1^k,\ldots,\ddot{m}_{M_s}^k\}$. If $M_s=M$, all APs participate into the transmission, without any selection. If $M_s=1$, only a single AP with the largest large-scale fading is determined, i.e.,
    \begin{equation}
        \ddot{m}_1^k=\arg \max_{m=1,\ldots,M} \bigl(\mathbf{b}_k\bigr),
    \end{equation} where $\mathbf{b}_k$ denotes the $k^{th}$ row of $\mathbf{B}$.
    \item \textit{Uplink Transmission:} User $k$, $\forall k$ transmits its data and particular pilot sequence over its assigned subcarriers $\mathbb{B}_k$. The opportunistic APs within  $\mathbb{M}_k$ estimate the uplink CSI $\hat{g}_{mk, n}$, where $m\in\mathbb{M}_k$ and $n\in \mathbb{B}_k$. The APs detect the uplink data coherently with the knowledge of uplink CSI.
    \item \textit{Conjugate Beamforming:} Afterwards, AP $m$, $\forall m\in \mathbb{M}_k$ knows the downlink CSI $\hat{g}_{mk, n}$ according to channel reciprocity. Then, it transmits the modulated symbol $s_{k,n}$, $n\in \mathbb{B}_k$ with $\mathbb{E}[|s_{k,n}|^2]=1$, and downlink pilot sequences over the assigned subcarriers $\mathbb{B}_k$. Applying conjugate beamforming in the frequency domain, the transmitted symbol at the $m^{th}$ AP is
   \begin{equation} \label{eqn:Txsignal}
    \tilde{x}_{m,n} = \sqrt{\eta_{mk}P_d} \hat{g}_{mk,n}^* s_{k,n},
   \end{equation} where $\sqrt{\eta_{mk}}$, $0\leq \eta_{mk}\leq 1$ denotes the power-control coefficient, and $P_d$ is a unified power constraint of each AP.
   \item \textit{Coherent Detection:} User $k$ estimates the downlink CSI $\hat{g}_{mk, n}$, where $m\in \mathbb{M}_k$ and $n\in \mathbb{B}_k$,  and detects the downlink data coherently with the aid of $\hat{g}_{mk, n}$.
\end{itemize}

\section{Performance Analysis}
We analyze the performance of three different schemes to shed light on the gains of opportunistic selection and downlink channel estimation. First, the performance in terms of spectral efficiency (SE) for the conventional CFmMIMO-OFDM system without opportunistic AP selection, denoted by  \textit{Full AP}, is analyzed, as a benchmark for comparison. Second, the performance of the system that selects $M_s$ opportunistic APs (\textit{OAS}) among $M$ APs but does not insert downlink pilots is also derived. Finally, the SE of opportunistic AP selection with the insertion of downlink pilots, denoted by \textit{OAS-DP}, is analyzed.

As a baseline, we directly apply the conventional conjugate beamforming in CFmMIMO \cite{Ref_bjornson2020scalable, Ref_nayebi2017precoding, Ref_zeng2021pilot, Ref_buzzi2020usercentric, Ref_jiang2021impactcellfree, Ref_masoumi2020performance} over each subcarrier in CFmMIMO-OFDM. To do so, each AP multiplexes a total of $K$ symbols, i.e., $s_{k,n}$ intended to user $k$, $k=1,\ldots,K$, before transmission. With a power-control coefficient $\sqrt{\eta_{mk}}$, $0\leq \eta_{mk}\leq 1$, the transmitted signal of the $m^{th}$ AP at subcarrier $n$ is
\begin{equation} \label{Eqn_ConjugateBFTranmitSymbol}
    \tilde{x}_{m,n} = \sqrt{P_d} \sum_{k=1}^{K} \sqrt{\eta_{mk}} \hat{g}_{mk,n}^* s_{k,n},
\end{equation}
where $\hat{g}_{mk,n}$ denotes an estimate of $\tilde{g}_{mk,n}$, and $\hat{g}_{mk,n} = \tilde{g}_{mk,n}-\xi_{mk,n}$ with estimation error $\xi_{mk,n}$ raised by additive noise.
Applying the minimum mean-square error (MMSE) estimation, we gets $\hat{g}_{mk,n}\in \mathcal{CN}(0,\alpha_{mk})$ with $\alpha_{mk}=\frac{P_u\beta_{mk}^2}{P_u\beta_{mk} + \sigma_z^2}$, where $P_u$ is the uplink power constraint, in comparison with $\tilde{g}_{mk,n}\in \mathcal{CN}(0,\beta_{mk})$. The details of channel estimation process can refer to \cite{Ref_jiang2021cellfree}.

In the conventional CFmMIMO, there is no downlink pilot and channel estimation due to the prohibitive overhead of inserting pilots over a massive number of antennae. Consequently, each user is assumed to only have the knowledge of channel statistics $ \mathbb{E} \left [ \left \vert \hat{g}_{mk,n}\right \vert ^2\right]= \alpha_{mk}$, a.k.a. channel hardening, rather than the channel realization $\hat{g}_{mk,n}$.
Substituting (\ref{Eqn_ConjugateBFTranmitSymbol}) into (\ref{Eqn_OFDMDL}), we get the received signal at user $k$, as shown in \eqref{Eqn_OFDMdownlink} on the top of the next page.
\begin{figure*}[!t]
\setcounter{equation}{14}
\begin{IEEEeqnarray}{lll}
\label{Eqn_OFDMdownlink} \nonumber
   \tilde{y}_{k,n} &=&\sqrt{P_d} \sum_{m=1}^M \tilde{g}_{mk,n} \sum_{k'=1}^{K} \sqrt{\eta_{mk'}}  \hat{g}_{mk',n}^* s_{k',n}+\tilde{z}_{k,n} \\ \nonumber
   &=&\underbrace{\sqrt{P_d} \sum_{m=1}^M \sqrt{\eta_{mk}} \mathbb{E} \left [\left|\hat{g}_{mk,n}\right|^2\right] s_{k,n}}_{\text{Desired signal}}  + \underbrace{\sqrt{P_d} \sum_{m=1}^M \sqrt{\eta_{mk}}\left( \left|\hat{g}_{mk,n}\right|^2-\mathbb{E} \left [\left|\hat{g}_{mk,n}\right|^2\right]\right) s_{k,n}}_{\text{Error due to CSI statistics}}  \\&+&\underbrace{\sqrt{P_d}\sum_{m=1}^M \hat{g}_{mk,n} \sum_{k'\neq k}^{K}  \sqrt{\eta_{mk'}}  \hat{g}_{mk',n}^* s_{k',n} }_{\text{Inter-user interference}}+\underbrace{\sqrt{P_d}\sum_{m=1}^M \xi_{mk,n} \sum_{k'= 1}^{K}  \sqrt{\eta_{mk'}}  \hat{g}_{mk',n}^* s_{k',n}}_{\text{Channel-estimation error}}+\underbrace{\tilde{z}_{k,n}}_{\text{Noise}}.
\end{IEEEeqnarray}
\end{figure*}
\setcounter{equation}{13}
Applying the method in \textit{Theorem 1}  of \cite{Ref_ngo2017cellfree}, we can derive that the spectral efficiency of user $k$ on subcarrier $n$, $\forall n=0,1,\ldots,N-1$  is lower bounded by $\log_2\left(1+\gamma_{k}^{\langle n\rangle}\right)$ with
\begin{equation} \label{eqn:SNR_SAAP}
    \gamma_{k}^{\langle n\rangle}=  \frac{ \left(\sum_{m=1}^M \sqrt{\eta_{mk}} \alpha_{mk}  \right)^2}
    { \sum_{m=1}^M \beta_{mk} \sum_{k'=1}^{K}  \eta_{mk'} \alpha_{mk'}+\frac{1}{\gamma_t} },
\end{equation}
with the transmit SNR $\gamma_t=P_d/\sigma^2_z$.

The proposed scheme exploits the degree of freedom enabled by the frequency domain to assign different users to orthogonal resources. As a result, the inter-user interference vanishes since each OFDM subcarrier accommodates a single user. Therefore, substituting $K=1$ into \eqref{eqn:SNR_SAAP} yields the performance of the first scheme with \textit{Full AP} transmission, i.e.,
\begin{equation}
    \gamma_{k}^{\langle n\rangle}=  \frac{ \left(\sum_{m=1}^M \sqrt{\eta_{mk}} \alpha_{mk}  \right)^2}
    { \sum_{m=1}^M \beta_{mk}   \eta_{mk} \alpha_{mk}+\frac{1}{\gamma_t} }.\end{equation}
To shed light on the effect of opportunistic selection, we then investigate the performance of selecting $M_s$ APs without adding downlink pilots. That is, each user only have the knowledge of channel statistics rather than the channel realization.
Substituting (\ref{eqn:Txsignal}) into (\ref{Eqn_OFDMDL}) to  get the received signal at user $k$ on subcarrier $n\in \mathbb{B}_k$ as \setcounter{equation}{15}
\begin{IEEEeqnarray}{lll} \label{Eqn_OFDMdownlink222} \nonumber
   \tilde{y}_{k,n} &=&\sqrt{P_d} \sum_{m\in \mathbb{M}_k } \tilde{g}_{mk,n} \sqrt{\eta_{mk}}  \hat{g}_{mk,n}^* s_{k,n}+\tilde{z}_{k,n} \\
   &=&\underbrace{\sqrt{P_d} \sum_{m\in \mathbb{M}_k} \sqrt{\eta_{mk}} \mathbb{E} \left [\left|\hat{g}_{mk,n}\right|^2\right] s_{k,n}}_{\text{Desired signal}}  \\  \nonumber
   &+& \underbrace{\sqrt{P_d} \sum_{m\in \mathbb{M}_k} \sqrt{\eta_{mk}}\left( \left|\hat{g}_{mk,n}\right|^2-\mathbb{E} \left [\left|\hat{g}_{mk,n}\right|^2\right]\right) s_{k,n}}_{\text{Error due to CSI statistics}}  \\
   \nonumber &+&\underbrace{\sqrt{P_d}\sum_{m\in \mathbb{M}_k} \xi_{mk,n} \sqrt{\eta_{mk}}  \hat{g}_{mk,n}^* s_{k,n}}_{\text{Channel-estimation error}}
   +\underbrace{\tilde{z}_{k,n}}_{\text{Noise}}.
\end{IEEEeqnarray}
Note that the received signal of user $k$ on subcarrier $n\in \{\mathbb{B}-\mathbb{B}_k\}$ is $\tilde{y}_{k,n}=0$.
Likewise, we can derive that the spectral efficiency of user $k$ on subcarrier $n\in \mathbb{B}_k$  is lower bounded by $\log_2\left(1+\gamma_{k}^{\langle n\rangle}\right)$ with
\begin{equation}
    \gamma_{k}^{\langle n\rangle}=  \frac{ \left(\sum_{m\in \mathbb{M}_k} \sqrt{\eta_{mk}} \alpha_{mk}  \right)^2}
    { \sum_{m\in \mathbb{M}_k} \eta_{mk} \beta_{mk}   \alpha_{mk} +\frac{1}{\gamma_t}}.
\end{equation}

Thanks to the opportunistic AP selection, there are only a few number of active APs over each subcarrier in the proposed scheme, while other far APs are turned off. From the perspective of a typical subcarrier, it is a low-dimensional MISO system, where the overhead of inserting downlink pilots is acceptable. As a result, user $k$ obtains the estimated CSI $\hat{g}_{mk,n}$ rather than the channel statistics $ \alpha_{mk}$. Thus, the received signal at user $k$ in \eqref{Eqn_OFDMdownlink222} can be reformed as
\begin{IEEEeqnarray}{lll} \nonumber
   \tilde{y}_{k,n} &=&\sqrt{P_d} \sum_{m\in \mathbb{M}_k } \tilde{g}_{mk,n} \sqrt{\eta_{mk}}  \hat{g}_{mk,n}^* s_{k,n}+\tilde{z}_{k,n} \\
   &=&\underbrace{\sqrt{P_d} \sum_{m\in \mathbb{M}_k} \sqrt{\eta_{mk}} \left|\hat{g}_{mk,n}\right|^2 s_{k,n}}_{\text{Desired signal}}  \\  \nonumber &+&\underbrace{\sqrt{P_d}\sum_{m\in \mathbb{M}_k} \xi_{mk,n} \sqrt{\eta_{mk}}  \hat{g}_{mk,n}^* s_{k,n}}_{\text{Channel-Estimation error}}
   +\underbrace{\tilde{z}_{k,n}}_{\text{Noise}}.
\end{IEEEeqnarray}

Applying the coherent detection, we can derive that the spectral efficiency of user $k$ on subcarrier $n\in \mathbb{B}_k$  is expressed by $\log_2\left(1+\gamma_{k}^{\langle n\rangle}\right)$ with
\begin{equation}
    \gamma_{k}^{\langle n\rangle}=  \frac{ \left(\sum_{m\in \mathbb{M}_k} \sqrt{\eta_{mk}} \left|\hat{g}_{mk,n}\right|^2   \right)^2}
    { \sum_{m\in \mathbb{M}_k} \eta_{mk} (\beta_{mk}-\alpha_{mk})   \alpha_{mk}+\frac{1}{\gamma_t} }.
\end{equation}

\section{Numerical Results}
The performance of the proposed opportunistic AP selection in CFmMIMO-OFDM is evaluated in terms of spectral efficiency.
We consider a square area of $1\times 1\mathrm{km^2}$ where a total of $M=128$ APs serve $K$ users. Deducting a guard band of $1\mathrm{MHz}$ at both sides, the signal bandwidth of $B_w=20\mathrm{MHz}$ is divided into $N=1200$ subcarriers with inter-subcarrier spacing $\triangle f=15\mathrm{kHz}$. Without losing generality, we ignores the overhead of pilot symbols in our simulation. As expected, it is observed in the simulations that the number of users does not affect the achievable spectral efficiency if $K$ is less than the number of subcarriers $N$, so that each subcarrier is assigned to a single user and there is no inter-user interference.

Large-scale fading is frequency independent and keeps constant for a relatively long period. It is given by $\beta_{mk}=10^\frac{PL_{mk}+X_{mk}}{10}$ with path loss $PL_{mk}$ and shadowing fading $X_{mk}\sim \mathcal{N}(0,\sigma_{sd}^2)$, where $\sigma_{sd}=8\mathrm{dB}$ is usually used. The COST-Hata model  \cite{Ref_ngo2017cellfree} is applied for calculating the path loss:
\begin{equation}
    PL_{mk}= \begin{cases}
-L-35\log_{10}(d_{mk}), &  d_{mk}>d_1 \\
-L-10\log_{10}(d_1^{1.5}d_{mk}^2), &  d_0<d_{mk}\leq d_1 \\
-L-10\log_{10}(d_1^{1.5}d_0^2), &  d_{mk}\leq d_0
\end{cases},
\end{equation}
where the three-slope breakpoints  take values $d_0=10\mathrm{m}$ and $d_1=50\mathrm{m}$ while $L=140.72\mathrm{dB}$ in terms of
\begin{IEEEeqnarray}{ll}
 L=46.3&+33.9\log_{10}\left(f_c\right)-13.82\log_{10}\left(h_{AP}\right)\\ \nonumber
 &-\left[1.1\log_{10}(f_c)-0.7\right]h_{UE}+1.56\log_{10}\left(f_c\right)-0.8
\end{IEEEeqnarray}
with carrier frequency $f_c=1.9\mathrm{GHz}$, the height of AP antenna $h_{AP}=15\mathrm{m}$, and the height of user antenna $h_{UE}=1.65\mathrm{m}$.
We generate small-scale fading using 3GPP Extended Typical Urban (ETU) model specified by delay profile in $\mathrm{millisecond}$ of $\{0,	0.05,	0.12,	0.2,	0.23,	0.5,	1.6,	2.3,	5\}$ with relative power $\{-1,	-1,	-1,	0,	0,	0,	-3,	-5,	-7\}$.
The maximum transmit power of AP and UE are $P_d=0.2\mathrm{W}$ and $P_u=0.1\mathrm{W}$, respectively. The APs simply adopt the full-power transmission strategy, i.e.,  $\eta_{m}=\left(\sum_{k=1}^{K} \alpha_{mk} \right)^{-1}$, $\forall m$.  The white noise power density  is $-174\mathrm{dBm/Hz}$ with a noise figure of $9\mathrm{dB}$.

\begin{figure}[!bpht]
\centering
\includegraphics[width=0.325\textwidth]{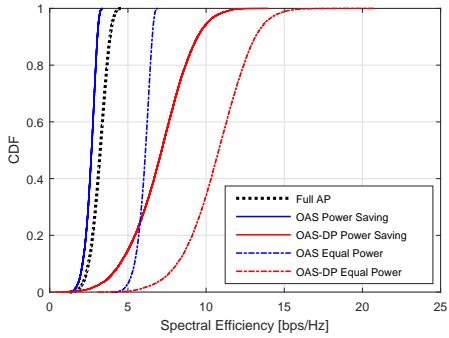}
\caption{CDFs of the achievable spectral efficiency for different schemes in a cell-free massive MIMO-OFDM system. }
\label{Diagram_Results}
\end{figure}

\figurename \ref{Diagram_Results} demonstrates the cumulative distribution functions (CDFs) of different schemes. First, the curve of \textit{Full AP} stands for the conventional CFmMIMO-OFDM system, where all $M=128$ APs serve an assigned user on a typical subcarrier without opportunistic AP selection.  The achieved $95\%$-likely spectral efficiency is around \SI{2.2}{\bps\per\hertz^{}} and the $50\%$-likely or median spectral efficiency is approximately \SI{3.2}{\bps\per\hertz^{}}. If $M_s=10$ active APs are selected in terms of large-scale fading, and the power constraint of each AP is the same as that of the \textit{Full AP}, the achieved SE of \textit{OAS Power Saving} is slightly inferior to the \textit{Full AP}. It has $95\%$-likely SE of around \SI{1.9}{\bps\per\hertz^{}} and the median SE of approximately \SI{2.7}{\bps\per\hertz^{}}. However, it significantly outperforms in terms of power efficiency since only $M_s=10$ APs are active, compared with $M=128$ APs in the \textit{Full AP}, amounting to a power saving of $92.19\%$. That is because that the power of the far APs cannot effectively transfer to the received power due to severe propagation loss. Turning off the far APs does not affect the total received power at the user.

As a fair comparison, we assume the selected APs has the same total power as the \textit{Full AP}, i.e., each opportunistic AP uses a power $M/M_s=12.8$ times higher. As shown by the CDF of \textit{OAS Equal Total Power}, the $95\%$-likely SE substantially increases to \SI{5.2}{\bps\per\hertz^{}} and the median SE reaches \SI{6.1}{\bps\per\hertz^{}}. Next, we can observe the significant performance gain of downlink pilots that enable the coherent detection at the user. Even if the total transmit power is less than $10\%$ of the \textit{Full AP}, \textit{OAS-DP Power Saving} achieves a $95\%$-likely SE of around \SI{3.8}{\bps\per\hertz^{}} and the median SE of \SI{7.2}{\bps\per\hertz^{}}. Compared with the \textit{Full AP}, it realizes the performance gain of about $70\%$ and $125\%$ in $95\%$-likely and median SE, respectively, while achieving a 10-fold power efficiency. Under the same total power, the superiority of opportunistic AP selection with the aid of downlink pilot is more significant. In this case, the $95\%$-likely SE substantially increases to \SI{7.4}{\bps\per\hertz^{}} and the median SE reaches \SI{10.8}{\bps\per\hertz^{}}. 

\section{Conclusions}
In this paper, we proposed to opportunistically select the best APs in cell-free massive MIMO-OFDM systems, taking advantage of \textit{the near-far} effect among different APs due to the cell-free structure. Exploiting the frequency domain enabled by OFDM transmission, users are assigned to orthogonal resource units (subcarriers or RBs).
By deactivating the far APs, which waste the energy over their channel with severe propagation loss, the transmission efficiency is improved. 
Numerical results corroborated that the proposed scheme can achieve a significant performance gain in terms of both spectral and power efficiencies.





%

\bibliographystyle{IEEEtran}
\bibliography{IEEEabrv,Ref_VTC2022}

\end{document}